\documentclass[usenatbib,usegraphicx]{mn2e}

\title[Dynamical state of the $\eta$ Cha sparse young open 
cluster]{Population and dynamical state of the $\eta$ Chamaeleontis
sparse young open cluster}

\author[A-R. Lyo et al.]
{A-Ran Lyo,$^{1}$\thanks{E-mail: arl@ph.adfa.edu.au (ARL);
wal@ph.adfa.edu.au (WAL); edf@astro.psu.edu (EDF);
lcrause@artemisia.ast.uct.ac.za (LAC)}
Warrick A. Lawson,$^{1\star}$
Eric D. Feigelson$^{1,2\star}$
and Lisa A. Crause$^{3\star}$\\
$^{1}$School of Physical, Environmental and Mathematical Sciences,
University of New South Wales,\\
$^{~}$Australian Defence Force Academy, Canberra ACT 2600, Australia\\
$^{2}$Department of Astronomy and Astrophysics, Pennsylvania
State University, University Park PA 16802, USA\\
$^{3}$Department of Astronomy, University of Cape Town,
Private Bag, Rondebosch 7700, South Africa}

\begin{document}

\date{Accepted for MNRAS}

\pagerange{\pageref{firstpage}--\pageref{lastpage}}
\pubyear{2003}

\maketitle

\label{firstpage}

\begin{abstract}

We report new results in our continuing study of the unique compact 
(1 pc extent), nearby ($d = 97$ pc), young ($t = 9$ Myr) stellar 
cluster dominated by the B9 star $\eta$ Chamaeleontis.  An optical 
photometric survey spanning $1.3 \times 1.3$ pc adds two M5 -- M5.5 
weak-lined T Tauri (WTT) stars to the cluster inventory which is 
likely to be significantly complete for primaries with masses 
$M > 0.15$ M$_\odot$.  The cluster now consists of 17 primaries and 
$\simeq 9$ secondaries lying within 100 AU of their primaries.  
The apparent distribution of 9:7:1 single:binary:triple systems 
shows $2-4 \times$ higher multiplicity than in the field main 
sequence stars, and is comparable to that seen in other pre-main 
sequence (PMS) populations.  The initial mass function (IMF) is 
consistent with that of rich young clusters and field stars.  By 
extending the cluster IMF to lower masses, we predict $10-14$ 
additional low mass stars with $0.08 < M < 0.15$ M$_\odot$ and 
$10-15$ brown dwarfs with $0.025 < M < 0.08$ M$_\odot$ remain to 
be discovered.  The $\eta$ Cha cluster extends the established 
stellar density and richness relationship for young open clusters.  
The radial distribution of stars is consistent with an isothermal 
sphere, but mass segregation is present with $> 50$ percent of 
the stellar mass residing in the inner 6 arcmin (0.17 pc).  
Considering that the $\eta$ Cha cluster is sparse, diffuse and 
young, the cluster may be an ideal laboratory for distinguishing 
between mass segregation that is primordial in nature, or arising 
from dynamical interaction processes.
      
\end{abstract}


\section{Introduction}

Most stars form in clustered environments, and it is now believed
that a significant fraction are born in sparse groups and clusters
with $N = 10-100$ stars \citep{Adams01, Elmegreen02}.  Study of the 
stellar and brown dwarf populations of sparse groups is therefore 
important for all star formation and stellar evolution issues, e.g.
the nature of the IMF, issues of dynamics and dispersal of young 
groups into the Galactic field, binarity, and the evolution of 
proto-planetary discs.

Our laboratory to study such issues is the $\eta$ Cha star 
cluster --- a recently-discovered sparse, young and nearby 
stellar group that is kinematically 
linked with the rich Sco-Cen (Sco OB2) Association (Mamajek, 
Lawson \& Feigelson 1999, 2000). Owing to its proximity to the 
Sun, compactness in the sky, and distance from obscuring clouds, 
the cluster provides an excellent laboratory for study of the
properties of $\sim 10$ Myr-old `intermediate-aged' PMS stars.

\citet{Mamajek99} discovered the $\eta$ Cha cluster from a deep
{\it ROSAT\,} High Resolution Imager observation. X-ray, proper
motion, photometric and spectroscopic study of the cluster had
resulted in the discovery of 15 primaries prior to this work
(see also Lawson et al. 2001, 2002).  In Sections 2 and 3, we 
update the cluster membership with the addition of two new 
M5 WTT stars, and we estimate the fraction of multiple systems 
in the cluster.  With a largely complete census for stars earlier 
than $\sim$ M6, we study the cluster IMF and its implications for 
the undiscovered low-mass star and brown dwarf population (Section 
4).  The spatial distribution of members provides evidence for 
mass segregation which has hithertofore been noted only in much 
richer clusters (Section 5).

\section{Observations and Data Reduction}

\subsection{Photometric selection of candidate members}

Optical colour-magnitude diagrams are a powerful tool aiding
the discovery and characterisation of PMS stellar populations
\citep{Walter00, Lawson01}.  For nearby compact, coeval and
codistant groups, PMS populations form a natural isochrone
in the colour-magnitude diagram that is elevated in brightness
above the location of a majority of field stars owing to a
combination of stellar youth, proximity and, for PMS populations
now distant from their parent molecular cloud, the absence of
significant interstellar reddening. For PMS groups discovered
via others means, e.g. X-ray study or objective prism surveys,
optical photometric study also permits an independent evaluation
of completeness within the survey field.  For example,
\citet{Lawson02} used the ($V-I$) versus $V$ colour-magnitude
diagram to locate X-ray faint PMS stars in the $\eta$ Cha cluster
that had similar photometric properties to the X-ray selected
RECX stars \citep{Mamajek99}.   In a limited area search of
$\approx 500$ arcmin$^{2}$ mostly near known members, two new 
cluster members were discovered including ECHA J0843.3--7905, 
a rare example of a classical T Tauri (CTT) star in an older 
PMS cluster or association.

\begin{figure*}
\begin{center}
\includegraphics[height=135mm]{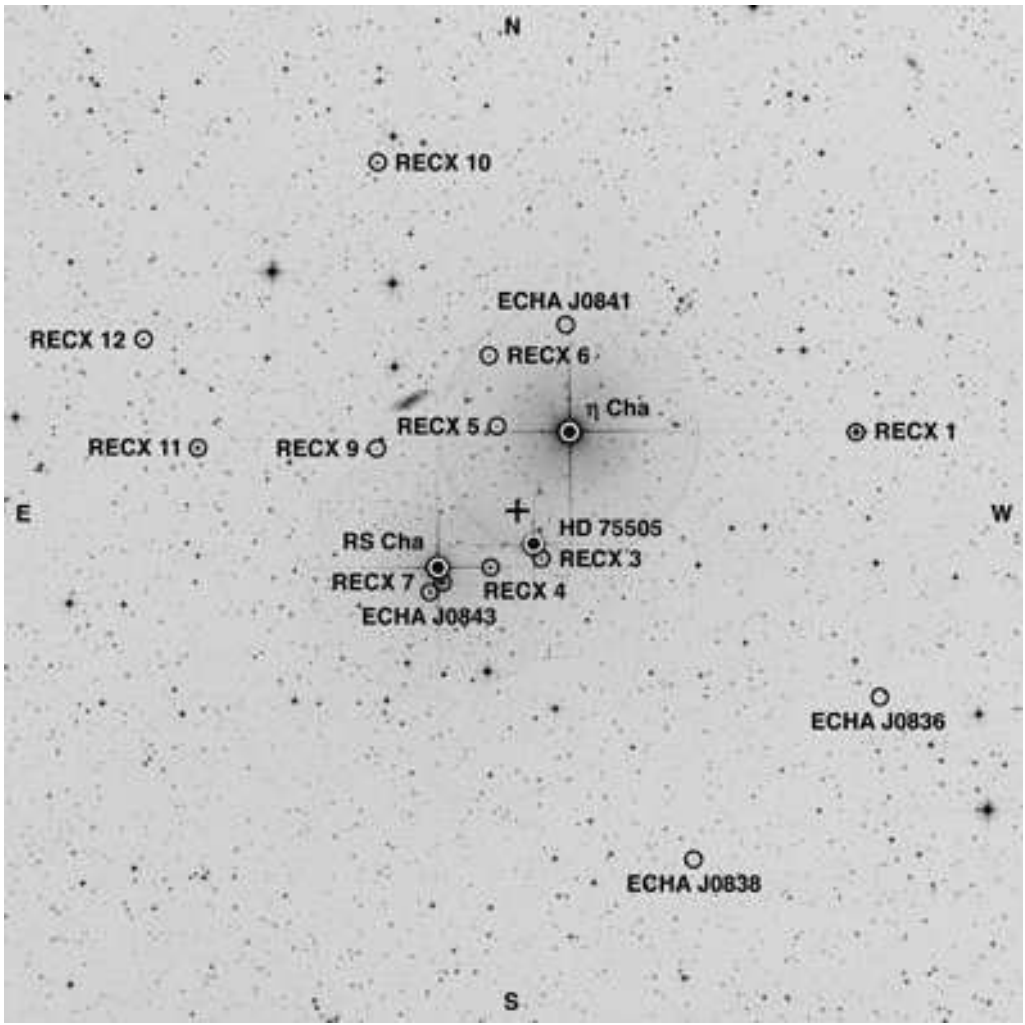}
\caption{$45 \times 45$ arcmin POSS-II red-band image of the 
$\eta$ Cha cluster region centered at $\alpha$, $\delta$ (2000) 
= 8h 42m 06s, $-79$d 01m 38s; the spatial centre of the 3 
early-type systems in the cluster (Mamajek et al. 2000).  The 
seventeen known cluster primaries are identified.}
\label{F1.fig}
\end{center}
\end{figure*}

In this paper we report the results of a photometric search of
the sky surrounding the $\eta$ Cha cluster, unbiased with respect
to location, spanning $45 \times 45$ arcmin ($\approx 2000$
arcmin$^{2}$, or $1.3 \times 1.3$ pc at $d = 97$ pc) and centred 
on $\alpha$, $\delta$ (2000) = 8h 42m 06s, $-79$d 01m 38s;
the spatial centre of the three early-type systems in the cluster 
\citep{Mamajek00}.  The extent of the survey field is shown in 
Figure \ref{F1.fig}, where cluster members are identified.

\begin{figure}
\begin{center}
\includegraphics[height=80mm]{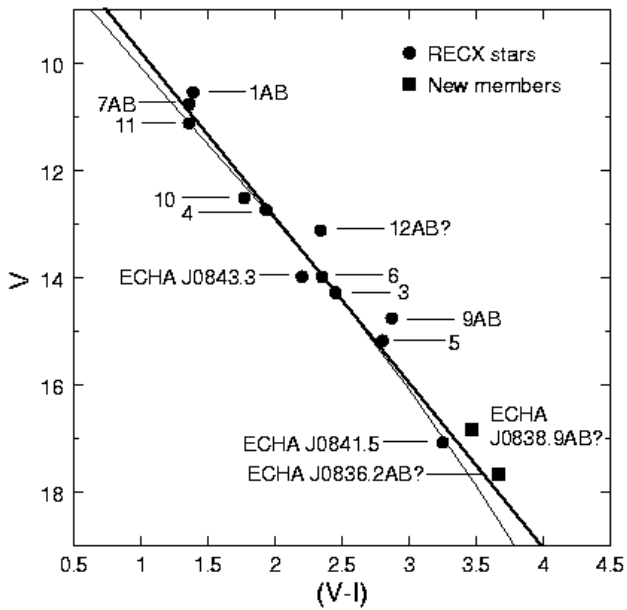}
\caption{Colour-magnitude diagram for the known late-type members
of the $\eta$ Cha cluster.  The bold line is the extrapolated linear
sequence of the late-type RECX stars from Lawson et al. (2001).
The curved line is a second-order approximation to the isochrone
for those stars likely to be single (or high-mass ratio binaries)
and those without photometry distorted by significant continuum
and line emission; see the text for details.  The RECX stars are 
denoted by number, with known and suspected binaries indicated.  
The ECHA-numbered stars are identified using abbreviated labels; 
see Table 2 for their full identifications.}
\label{CMD.fig}
\end{center}
\end{figure}

We used the 1-m telescope and 1k $\times$ 1k SITe charge-coupled
device (CCD) at the Sutherland site of the South African 
Astronomical Observatory (SAAO) in 2002 January to map the 
field in the Cousins $V$ and $I$ bands.
The data were obtained under photometric conditions in $1.5-3$
arcsec seeing, with the raw aperture-extracted observations of
stars in each CCD field then transformed to the standard system
using observations of southern photometric standard stars and
extinction values measured for the Sutherland site.  Exposure
times in the $V$ and $I$ bands of 90 s and 30 s, respectively,
permitted detection (for $\sim 10$ Myr-old PMS stars located at
$d \approx 100$ pc) of stars of spectral type $\approx$ M6 with
$V \approx 18$ and ($V-I$) $\approx 4.5$ at a photometric
precision better than 3 percent.  Following \citet{Lawson02},
we investigated stars with properties approximate to known members,
choosing objects with magnitudes in the range of $-1.0 < V < 1.0$
mag of the sequence of RECX stars in the ($V-I$) versus $V$
colour-magnitude diagram.  This criterion is intended to account
for the possibility of a greater spread of ages within the cluster
than is apparent in the RECX stars, highly reddened stars or
binary stars with elevated $V$ magnitudes, and the likelihood
of non-linearities in the colour-magnitude relationship for the
cluster \citep{LawsonFeigelson01}.   With the exception of the
three bright early-type stars and the two brightest K-type 
members which saturated the CCD, the late-type clusters members 
were recovered in the survey along with several of the 
photometrically-similar non-members discussed by \citet{Lawson02}.  
Figure \ref{CMD.fig} shows the photometric sequence for known and 
candidate late-type members of the cluster. The bold line in this 
diagram is the extrapolated linear sequence of the RECX stars 
\citep{Lawson01}.  This line likely represents a first-order 
approximation of the real shape of the cluster isochrone in the 
($V-I$) versus $V$ colour-magnitude plane, but has proven useful 
for identifying candidate cluster members.  Other details of this 
diagram are discussed below.

The survey resulted in the identification of two new candidate 
members, and we detail their spectroscopic characterisation in 
the following sections.

\subsection{Spectroscopic confirmation}

Medium-resolution optical spectroscopy of the two new candidates was 
obtained during 2002 March using the 2.3-m telescope and dual-beam 
spectrograph (DBS) at Mount Stromlo and Siding Spring Observatories 
(MSSSO).  In the red beam, the 1200R (1200 line\,mm$^{-1}$) grating 
gave a 2-pixel resolution of 1.1 \AA\, with coverage from 
$\lambda\lambda 6200-7160$ \AA.  Exposure times of $3 \times 1200$ 
s for these stars yielded continuum signal-to-noise (S/N) ratios of 
$15-20$ near H$\alpha$. The spectra were calibrated using dome-flats, 
bias frames and Ne-Ar arc frames, making use of standard library 
routines such as {\tt ccdproc} within {\tt IRAF}.

\begin{figure}
\begin{center}
\includegraphics[width=80mm]{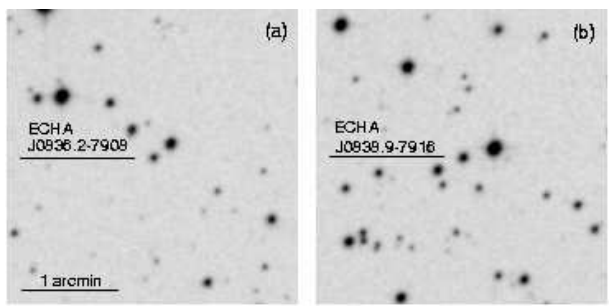}
\caption{POSS-II red-band finder charts (width = 3 arcmin) 
centred on the positions of (a) ECHA J0836.2--7908 (b) ECHA 
J0838.9--7916.}
\label{Finder.fig}
\end{center}
\end{figure}

Analysis of the spectra showed that both of these stars were active, 
lithium-rich late-type objects.  These two stars (henceforth 
ECHA J0838.9--7916 and ECHA J0836.2--7908, respectively) are assigned 
an $\eta$ Cha cluster designation making use of J2000 coordinates.
Identification fields, derived from the Second Palomar Observatory 
Sky Survey (POSS-II), are shown in Figure \ref{Finder.fig}.  Both 
stars reside outside the {\it ROSAT\,} HRI discovery observation 
of \citet{Mamajek99} and are located $17-18$ arcmin SW of the 
cluster centre.

To accurately spectral type the two new cluster members, we also 
obtained low-resolution 2.3-m/DBS spectra (2 \AA-resolution in 
the blue arm, and 4 \AA-resolution in the red arm) using the 300B 
(300 line\,mm$^{-1}$) and 158R (158 line\,mm$^{-1}$) gratings during 
2002 March.  The blue and red spectra were obtained simultaneously 
in a pseudo-spectrophotometric mode; with the slit width set to 
maximise the spectral resolution, and oriented vertically to 
eliminate differential refraction.  The spectra were calibrated 
using flat-spectrum sources and flux calibrators, and when combined 
had continuous coverage from $\lambda\lambda 3500-10200$ \AA.   
These data were compared to spectra of M-type dwarfs from the 
LHS catalogue obtained with the same instrumentation; see 
\citet{Bessell90, Bessell91} for a list of late-M LHS stars, and
\citet{Bessell99} for a description of the spectroscopic reduction 
techniques used here. Spectra obtained in this manner permit comparative
spectral typing and extraction of synthetic colours, but not the 
measurement of fluxes.

\section{Cluster membership}

The stellar population of the $\eta$ Cha cluster has been
progressively defined via several methods; X-ray detection and 
astrometric study \citep{Mamajek99, Mamajek00} and photometric 
selection \citep{Lawson02}, with subsequent spectroscopic study
confirming stellar youth in late-type members. \citet{Mamajek99} 
characterised twelve X-ray active stars detected using {\it ROSAT\,}
HRI; $\eta$ Cha, RS Cha and ten late-type members ranging from 
spectral types K4 -- M5, and added the X-ray quiet HD 75505 with 
Tycho astrometry consistent with the {\it Hipparcos\,} distances 
and proper motions of $\eta$ Cha and RS Cha.  From optical
photometric selection, \citet{Lawson02} added two X-ray faint 
stars residing within the {\it ROSAT\,} HRI field, including the 
CTT star ECHA J0843.3-7905.  In this paper we complete an extended 
photometric survey with the aim of detecting all late-type cluster 
primaries within $r = 23-32$ arcmin of the cluster centre, and 
announce the discovery of two new low-mass members which likely
completes the inventory of cluster primaries within this region.

\subsection{New members}

\subsubsection{ECHA J0836.2--7908}

The DBS H$\alpha$ region spectrum of ECHA J0836.2--7908 (Figure
\ref{DBS.fig}) shows moderate levels of H$\alpha$ emission for 
a M-type PMS star (equivalent width $EW = -8.3$ \AA) and strong
$\lambda 6707$ Li I absorption ($EW = 0.6$ \AA) comparable to other 
$\eta$ Cha cluster late-type members \citep{Mamajek99, Lawson02}.  

For ECHA J0836.2--7908 we determined a spectral type of $\approx$ 
M5.5 from the low resolution DBS spectra.  The star has optical 
and 2MASS photometry consistent with a M5.5 spectral classification, 
e.g. ($V-I$) = 3.76 
(M5.5), ($V-K$) = 6.71 (M5.5) and ($I-K$) = 2.95 (M5.5) using data 
presented in Table \ref{T1.tab} and employing the spectral type-colour
conversions of \citet{Bessell91}.  With $V = 17.66$ and an inferred 
mass of $M \approx 0.15$ M$_{\odot}$ using Siess, Dufour \& Forestini 
(2000) tracks, this star is the faintest and lowest-mass primary known 
in the cluster.  The position of the star in the ($V-I$) versus $V$ 
colour-mag diagram (Figure \ref{CMD.fig}) is suggestive of the star 
being binary, which we discuss in Section 3.2.1.

Multi-epoch {\it VRI\,} CCD photometry was obtained with the 1-m
telescope at SAAO during 2002 April.  22 epochs were obtained over
a 12 d interval, with exposure times of 90 s, 60 s and 30 s in the 
{\it VRI\,} bands, respectively.  The reduction and analysis of 
the multi-epoch data followed that described by \citet{Lawson01}.
Fourier analysis showed no significant periodicities were present
in the light curves.  The {\it VRI\,} data sets showed scatter at 
the 1-$\sigma$ level of 0.03 mag (in $V$), 0.01 mag (in $R$) and 
0.01 mag (in $I$); levels similar to comparably-bright field stars.

\subsubsection{ECHA J0838.9--7916}

\begin{figure}
\begin{center}
\includegraphics[width=80mm]{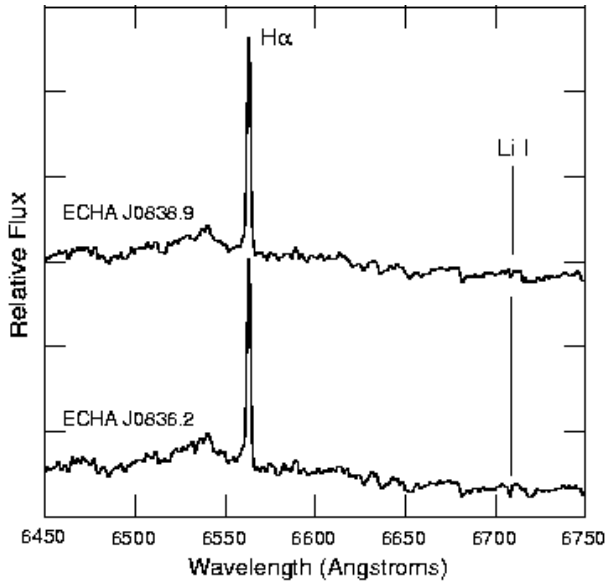}
\caption{H$\alpha$ region spectroscopy of ECHA J0836.2--7908 and 
ECHA J0838.9--7916 showing H$\alpha$ emission and $\lambda 6707$
Li I absorption lines.}
\label{DBS.fig}
\end{center}
\end{figure}

H$\alpha$ region DBS spectroscopy of ECHA J0838.9--7916 (Figure 
\ref{DBS.fig}) shows moderate H$\alpha$ emission ($EW = -8.0$ \AA) 
and $\lambda 6707$ Li I absorption ($EW = 0.3$ \AA) lines.   
Low-resolution DBS spectroscopy indicates a spectral type of
M5 -- M5.5, which we confirm from the measured photometric colours,
e.g. ($V-I$) = 3.54 (M5.2), ($V-K$) = 6.39 (M5.2), and ($I-K$) = 
2.85 (M5.2); see Table \ref{T1.tab}.   We adopt a spectral type 
of M5.2 for this star.  From comparison with the evolutionary tracks 
of \citet{Siess00} we infer a mass of $M \approx 0.16$ M$_{\odot}$.
Like ECHA J0836.2--7908, the position of ECHA J0838.9--7916 in the
($V-I$) versus $V$ colour-mag diagram (Figure \ref{CMD.fig}) is 
suggestive of the star being binary; see Section 3.2.1.

Multi-epoch (21 epochs) {\it VRI\,} photometry obtained at SAAO
during 2002 April showed no evidence for periodicity.  The {\it VRI\,} 
data sets showed scatter at the 1-$\sigma$ level of 0.01 mag in all
three photometric bands; levels similar to comparably-bright field 
stars.

The lack of variability (see also the previous section on ECHA 
J0836.2--7908) is surprising when periodic behaviour due to the 
rotational modulation of cool starspots or accretion hotspots 
has been detected in all other late-type cluster members which
span spectral types K5 -- M4 \citep{Lawson01, Lawson02}.  
Either the light curves of both these new M5 members are 
undersampled, or both stars have their rotation axes aligned
pole-on, or we have detected a significant reduction in magnetic 
activity in these $\sim 10$ Myr-old M5 PMS stars.  The photometric 
noise level of our observations indicates the spot coverage of 
both stars (strictly, the difference in coverage during the star's 
rotation period) cannot exceed $1-2$ percent.  This is a factor 
of $2-4$ lower than the light amplitudes observed in the M4 cluster 
members RECX 5 and ECHA J0841.5--7853.

The USNO--B1.0 catalogue lists a proper motion for ECHA J0838.9--7916 
of ($\mu_{\alpha}$, $\mu_{\delta}$) = ($-34 \pm 3$, $24 \pm 12$)
mas\,yr$^{-1}$.  The proper motion is consistent with the weighted 
mean of Tycho-2 proper motions for the brightest cluster members of 
($\mu_{\alpha}$, $\mu_{\delta}$) = ($-30.0 \pm 0.3$, $27.8 \pm 0.3$) 
mas\,yr$^{-1}$ \citep{Mamajek00}.

\begin{table*}
\centering
\caption{J2000 position, Cousins {\it VRI\,} and 2MASS {\it JHK\,}
photometry, and spectroscopic features of the two new members of 
the $\eta$ Cha cluster.   The spectral types have an estimated 
uncertainty of 0.1 subtype; see Section 3.1 for details.} 
\label{T1.tab}
\begin{tabular}{@{}lcccccccccccc@{}}
\hline
 & $\alpha_{2000}$ & $\delta_{2000}$ & & & & & & 
 & $H_{\alpha}~EW$ & Li I~$EW$ & Spectral\\
Star & (h~m~s) & (d~m~s) & $V$ & $R$ & $I$ & $J$ & $H$ & $K$ 
 & (\AA) & (\AA) & type\\ \hline
ECHA J0836.2--7908 & 08 36 10.6 & --79 08 18 & 17.66
 & 15.94 & 13.90 & 11.85 & 11.28 & 10.95 & $-8.3$ & 0.6 & M5.5 \\
ECHA J0838.9--7916 & 08 38 51.5 & --79 16 14 & 16.82
 & 15.21 & 13.28 & 11.28 & 10.72 & 10.43 & $-8.0$ & 0.3 & M5.2 \\
\hline
\end{tabular}
\end{table*}

\subsection{An improved census of the $\eta$ Cha cluster}

With no new cluster members detected between $r = 18 - 23$ arcmin
distance from the cluster centre, and with coverage as distant 
as $r = 32$ arcmin in the corners of the $45 \times 45$ arcmin 
surveyed region (Figure \ref{F1.fig}), our X-ray, astrometric 
and photometric surveys has likely discovered the full extent of 
the cluster's stellar primary population with spectral types 
earlier than $\sim$ M6, unless other stars have been distantly
removed by evaporation or dynamical ejection.

Table \ref{T2.tab} lists the known stellar primaries associated 
with the cluster.  A number of these stars are in known or 
suspected (indicated by the `?' in the column of companion 
masses in Table \ref{T2.tab}) multiple systems.   Other than for 
RS Cha AB which has accurate physical masses, we infer the mass 
of the primary stars from comparison with the PMS evolutionary
tracks of \citet{Siess00}.  The presence of companions and
their masses are inferred using a range of evidence that we 
detail in the following sub-sections.  For the late-type stars
our current observations are sensitive only to low mass ratio 
binaries, thus the frequency of systems with higher mass ratios 
is unknown and therefore the level of multiplicity is likely 
to be higher than we discuss here (see also Section 4 for 
discussion of the cluster IMF).

\subsubsection{ECHA J0836.2--7908 and ECHA J0838.9--7916}

Compared to the location of ECHA J0841.5-7853 in the ($V-I$)
versus $V$ colour-magnitude diagram (Figure \ref{CMD.fig})
both these new cluster members appear to have elevated $V$
magnitudes for their ($V-I$) colour.  We illustrate this point
in Figure \ref{CMD.fig} by fitted a second-order polynomial 
to those stars (see Section 3.2.7) we believe are single or
high mass ratio binaries (excluding the CTT star ECHA
J0843.3-7905 with photometry affected by continuum and/or line
emission).  This curve probably better represents the isochrone 
for the low-mass members than the linear fit of \citet{Lawson01}, 
also shown in Figure \ref{CMD.fig}.  If ECHA J0841.5--7853 defines 
the location of the cluster isochrone for late-M spectral types, 
then both ECHA J0836.2--7908 and ECHA J0838.9--7916 appear
over-luminous by $\approx 0.7$ mag and are candidate 
near-equal mass binary systems.

\subsubsection{RECX 1 and RECX 9}

Both stars are elevated by $0.5-0.7$ mag in the ($V-I$) versus 
$V$ colour-magnitude diagram compared to other RECX stars of 
similar spectral type (see Figure \ref{CMD.fig}) suggesting 
near-equal mass binary systems.  Speckle $K$-band imaging by
\citet{Kohler02} using the European Southern Observatory 3.5-m 
NTT found both stars have companions at separations of $0.1-0.2$ 
arcsec ($10-20$ AU projected separation at  $d = 97$ pc).  
RECX 1AB and RECX 9AB have $K$-band brightness ratios of 
$\approx 0.8$ and 0.5, respectively.  RECX 1A is a mid-K star, 
indicating RECX 1B is a late-K star.  RECX 9A has a spectral 
type of M4, indicating RECX 9B is a M5 star.

\subsubsection{RECX 2 = $\eta$ Cha}

$\eta$ Cha was detected in X-rays by {\it ROSAT\,} HRI with log
$L_{X} = 28.8$ erg\,s$^{-1}$ \citep{Mamajek00}.  The X-ray emission 
is likely associated with a low mass companion.  Employing the broad
$L_{X}-M$ relationship seen in Orion PMS stars \citep{Feigelson03},
we tentatively assign the companion to be an early-M star with 
$M = 0.50$ M$_{\odot}$.

\subsubsection{RECX 7}

RECX 7 is a dual-lined spectroscopic binary with a mid-K spectral 
type primary and a binary mass-ratio of $\approx 2.3:1$ \citep{Lyo03}.
If RECX 7A is a near-solar mass star, then RECX 7B is a $\sim 0.5$ 
M$_{\odot}$ early-M star.

\subsubsection{RECX 8 = RS Cha AB}

RS Cha AB is a A7+A8 dual-lined eclipsing system with accurate 
physical masses of $M_{A} = 1.86$ M$_{\odot}$ and $M_{B} = 1.82$ 
M$_{\odot}$ \citep{Andersen91, Mamajek00}.  RS Cha was detected 
in X-rays by {\it ROSAT\,} HRI with log $L_{X} = 29.9$ erg\,s$^{-1}$
\citep{Mamajek00}.  As for $\eta$ Cha, we tentatively associate 
the X-ray emission with a low mass ($M \approx 0.5$ M$_{\odot}$ 
tertiary) companion.

\subsubsection{RECX 12}

RECX 12 is elevated by $\approx 0.7$ mag in Figure \ref{CMD.fig}
compared to other cluster stars of similar spectral type,
suggesting the star is a near-equal mass binary.  Other 
evidence for binarity includes the detection of dual periodicities
in the light curve of the star measured in 1999 and 2000.  
\citet{Lawson01} detected 1.3 d and 8.6 d periods in a 
multi-epoch $V$ band study in both years.  Also, \citet{Kohler02} 
suggested RECX 12 may be a binary, but unresolved in their
$K$-band speckle observations.

\subsubsection{Other cluster members}

Our observations suggest ECHA J0841.5--7853, RECX 3, 
HD 75505, RECX 4, RECX 5, RECX 6, ECHA J0843.3--7905, RECX 10 
and RECX 11 are likely single stars, or possibly high-mass
ratio binary systems.

\begin{table}
\centering
\caption{Membership of the $\eta$ Cha cluster,
ordered in increasing Right Ascension.  The radius of each system 
from the cluster centre is given.  Primary and compansion masses
are physical masses (for RS Cha AB) or inferred from comparison
with the evolutionary grids of Siess et al. (2000).  Probable
companions are indicated with a `?'.  Chamaeleon variable star 
designations for the low-mass RECX stars are from Kazarovets 
et al. (2003).}
\label{T2.tab}
\begin{tabular}{@{}lccc@{}}
\hline
Star &  Radius      & Primaries    & Companions    \\
     & (arcmin)     & (M$_{\odot}$)&(M$_{\odot}$)  \\ \hline

ECHA J0836.2--7908  & 18.02 & 0.15 & 0.15?      \\
RECX 1 (EG Cha)     & 15.63 & 1.00 & 0.70       \\
ECHA J0838.9--7916  & 17.18 & 0.16 & 0.16?      \\
RECX 2 ($\eta$ Cha) &  4.43 & 3.40 & 0.50?      \\
ECHA J0841.5--7853  &  8.70 & 0.19 & ---        \\
RECX 3 (EH Cha)     &  2.33 & 0.32 & ---        \\
HD 75505            &  1.61 & 1.85 & ---        \\
RECX 4 (EI Cha)     &  2.56 & 0.49 & ---        \\
RECX 5 (EK Cha)     &  3.96 & 0.26 & ---        \\
RECX 6 (EL Cha)     &  7.09 & 0.35 & ---        \\
RECX 7 (EM Cha)     &  4.35 & 1.08 & 0.50       \\
RECX 8 (RS Cha)     &  4.06 & 1.86 & 1.82+0.50? \\
ECHA J0843.3--7905  &  5.04 & 0.37 & ---        \\
RECX 9 (EN Cha)     &  6.71 & 0.20 & 0.18       \\
RECX 10 (EO Cha)    & 16.52 & 0.60 & ---        \\
RECX 11 (EP Cha)    & 14.27 & 1.04 & ---        \\
RECX 12 (EQ Cha)    & 18.16 & 0.35 & 0.35?      \\ \hline

\end{tabular}
\end{table}

\section{Mass function of the $\eta$ Cha cluster}

The stellar initial mass function (IMF) is important for studies 
of star formation and stellar evolution in clusters and galaxies. 
Surprisingly, the IMF has been shown to be essentially uniform 
in space and time, suggesting that the basic process of the star 
formation is somewhat universal \citep{Elmegreen02}.  However, 
IMF studies are usually confined to rich clusters or widespread 
field populations, and have rarely been examined in a sparse 
cluster environment.

\subsection{Comparison with rich young clusters}

Using the membership and masses in Table \ref{T2.tab}, we compare
the IMFs of the sparse $\eta$ Cha cluster and rich Orion Trapezium
Cluster in Figure 5.  For the Trapezium cluster, we adopt power-law
slopes $\Gamma = -2.21$  for $M_{*} > 0.600$ M$_{\odot}$, $-1.15$
for $0.600 > M_{*} > 0.120$ M$_{\odot}$, and $-0.27$ for $0.120
> M_{*} > 0.025$ M$_{\odot}$, where $\Gamma$ = $d$\,log $N$($N_{*} 
>$ log $M_{*}$)/$d$\,log $M_{*}$ \citep{Muench02}.

For both the distributions of primary (Figure 5a) and member masses 
(Figure 5b), we scale the Trapezium IMF to the observed cumulative 
distributions for $\eta$ Cha by minimizing the Kolmogorov-Smirnov 
(K-S) statistic, or supremum difference between the distributions.  
We find the distribution of primaries/members is consistent with the 
Trapezium IMF at the $P = 90$ percent confidence level according to 
the K-S test, although we suspect a few low mass stars may remain 
to be discovered in high-mass ratio multiple systems.  Also, if 
the cluster population does significantly extend beyond the limits 
of the searched area, then they are likely to be low-mass members 
ejected by dynamical interactions.  A number of recent dynamical 
studies \citep{Sterzik01, Delgado03} find small stellar groups can 
evolve toward evaporation and dispersal on timescales of a few Myr, 
with an outer halo dominated by low-mass single stars.

Extending the comparison to masses $M < 0.15$ M$_{\odot}$, we 
predict (for primaries; Figure 5a) an additional 10 stars between 
$0.08 < M < 0.15$ M$_{\odot}$ and 10 brown dwarfs between $0.025 
< M < 0.08$ M$_{\odot}$.  For all members (primaries and companions) 
the comparable numbers are 14 and 15, respectively.  Thus from this 
calculation we expect the undiscovered low-mass population (objects 
with $M > 25$ M$_{J}$) of the $\eta$ Cha cluster is comparable in 
number to the known stellar population, if this cluster follows 
the IMF of a rich young stellar clusters such as Trapezium.

\subsection{Comparison with sparse young groups}

We also consider the suggestion of \citet{Adams01} that young 
small groups and rich clusters represent different types of 
physical systems and hence different models of star formation.  
Therefore, we compare the mass function of $\eta$ Cha cluster 
with the mass function of other young small groups.  In this 
case, we use the ratio $\Re$ of intermediate- to low-mass stars 
where $\Re$ = $N$($1-10$ M$_{\odot}$)/$N$($0.1-1$ M$_{\odot}$) 
because the mass bins are wider than the uncertainties in stellar 
mass determination from PMS models and to alleviate the statistical 
issues raised when comparing mass functions of groups with small
populations.   For the $\eta$ Cha cluster we obtain $\Re$ = 
$0.55 \pm 0.39$ and $0.37 \pm 0.22$ for primaries and members, 
respectively.  These values are consistent with the range of 
ratios $\Re = 0.07-0.45$ found for nine small young groups/clusters 
containing 27 to 133 members (see Table I of Meyer et al. 2000).

\subsection{Comparison with the field IMF}

The observed ratios for $\eta$ Cha are consistent with the 
expected ratios (for primaries) using the IMF's of \citet{Miller79}
($\Re = 0.185$) and \citet{Kroupa01} ($\Re = 0.199$). 

Merging the above comparisons, we find no significant difference 
between the IMF for the $\eta$ Cha cluster, when compared to other
small young stellar groups, rich young clusters, or field stars.

\begin{figure*}
\begin{center}
\includegraphics[height=80mm]{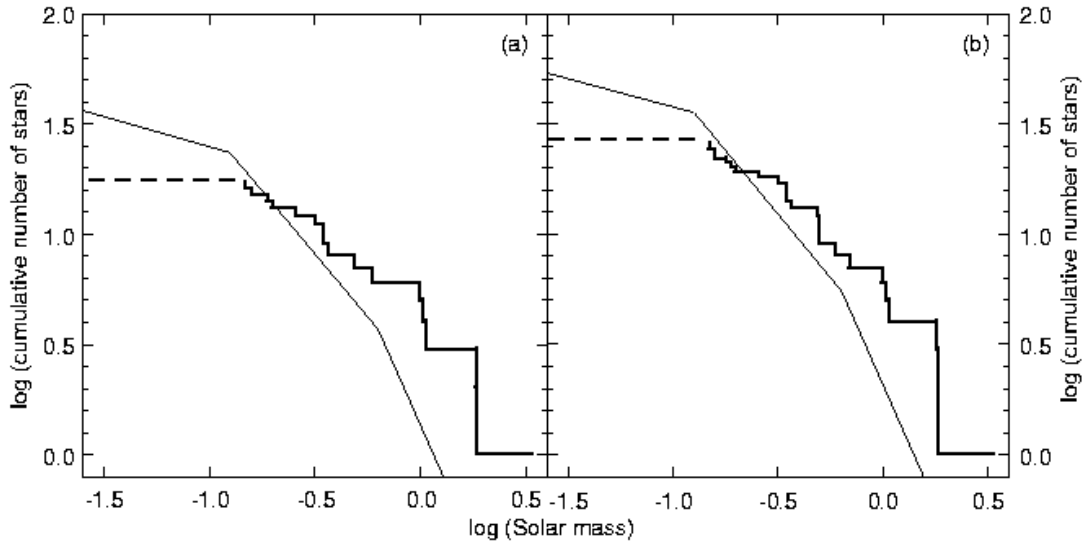}
\caption{Mass function for the $\eta$ Cha cluster for (a) primaries
and (b) members including known and probable companions.  In both 
(a) and (b) we compare the cluster mass function to the Trapezium 
IMF (thin lines); see Section 4.1 for details.}
\label{F5.fig}
\end{center}
\end{figure*}

\section{Population and mass distributions of the $\eta$
Cha cluster}

\subsection{Mass segregation and dynamical state}

\begin{figure}
\begin{center}
\includegraphics[width=80mm]{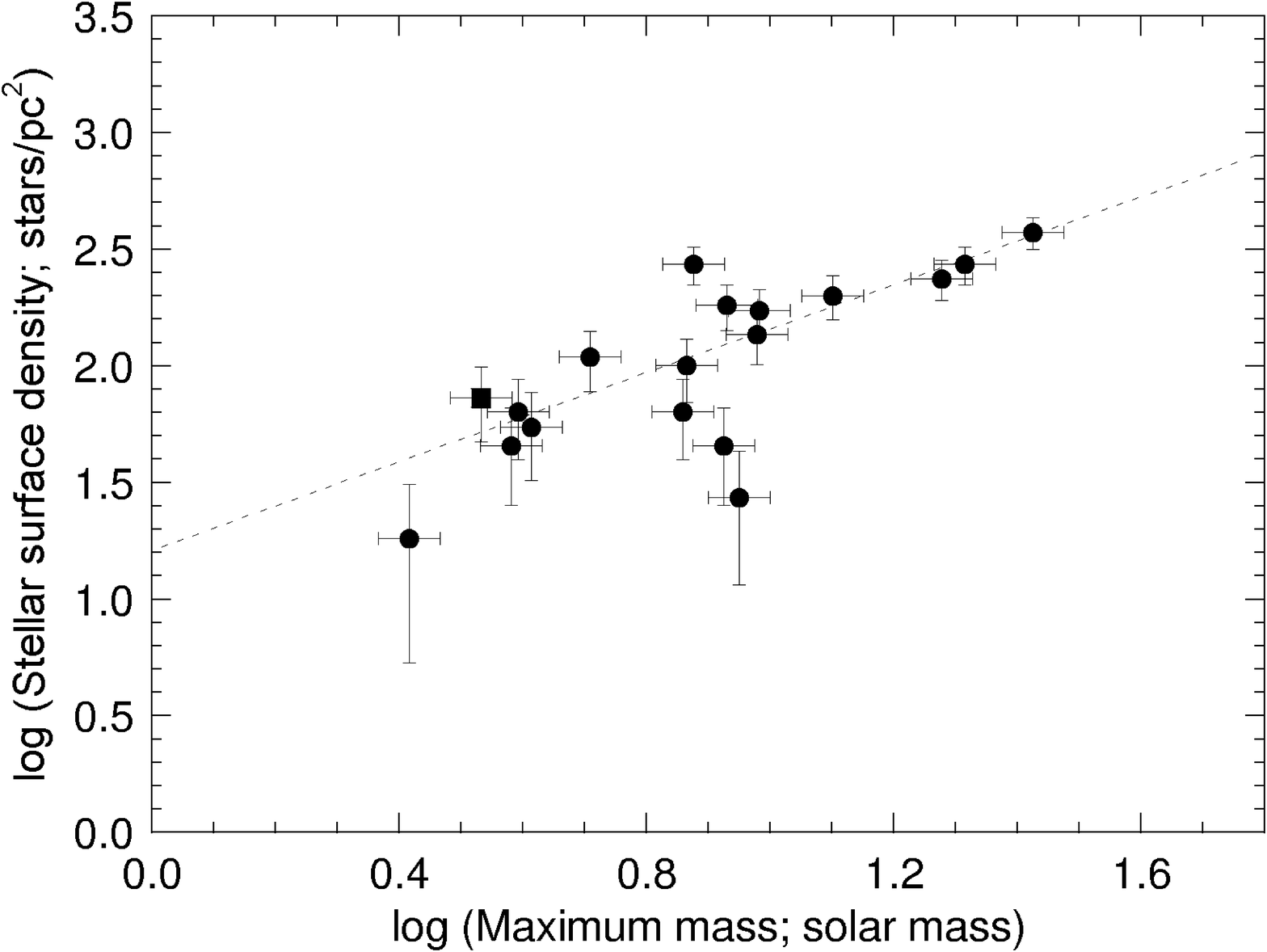}
\caption{Stellar surface density (stars\,pc$^{-2}$; measured within
the inner 0.33 pc) compared to the mass (M$_{\odot}$) of the most
massive cluster member.  Values other than for the $\eta$ Cha cluster 
(filled square) are adopted from Hillenbrand (1995).  Random errors 
of $\pm 5$ percent in mass and $\sqrt{N}$ in star counts are shown, 
along with a least-square fit to these data.  The comparison shows
the $\eta$ Cha cluster follows the `richness' -- mass pattern seen
in other compact groups and open clusters; see Section 5.1.}
\label{F6.fig}
\end{center}
\end{figure}

\citet{Hillenbrand95} shows a correlation between the stellar 
density of compact groups and open clusters, and the mass of the 
most massive cluster member; also see figure 2 of Clarke, Bonnell 
\& Hillenbrand (2000).   The $\eta$ Cha cluster follows the 
correlation seen in other clusters; see Figure \ref{F6.fig}.
Following \citet{Hillenbrand95}, we obtained the measure of cluster 
`richness' by counting the number of primaries in the central 0.33 pc 
($r = 6$ arcmin).  For the $\eta$ Cha cluster, we count 8 primaries
within 6 arcmin of our adopted cluster centre lead by the $M = 3.4$ 
M$_{\odot}$ B9 star $\eta$ Cha.  

Studies have also shown that the most massive stars of rich 
stellar clusters are concentrated in the central core, e.g. Mon R2 
\citep{Carpenter97}, the Orion Nebula Cluster \citep{Hillenbrand98}, 
NGC 6231 \citep{Raboud98} and NGC 2157 in the LMC \citep{Fischer98}.  
Employing $N$-body simulations of stellar cluster populations, 
\citet{Bonnell98} suggest that mass segregation in rich clusters
reflects the initial conditions of the cluster, and is not a 
consequence of dynamical interactions.  

\begin{figure}
\begin{center}
\includegraphics[width=84.5mm]{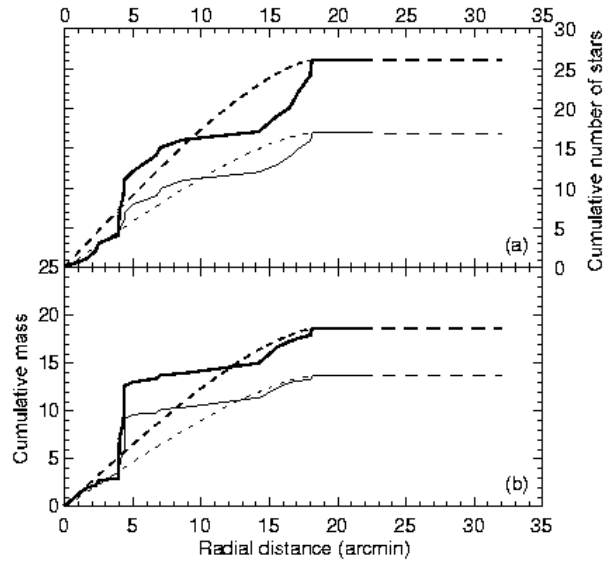}
\caption{Cumulative radial (a) number and (b) mass distribution 
for primaries (thin lines) and members (bold lines) in the $\eta$ 
Cha cluster.  These lines are horizontal for $r > 18$ arcmin from 
the cluster centre, reflecting our non-detection of cluster members
beyond this radius.  The dashed lines for $23 < r < 32$ arcmin 
reflect the partial completeness of our survey at these radii; 
see Section 2.  The dotted lines for $r < 18$ arcmin are the 
distributions for isothermal spheres; see Section 5.1 for a
statistical evaluation of these distributions.}
\label{F7.fig}
\end{center}
\end{figure}

In Figure \ref{F7.fig}, we plot the cumulative radial (a) number 
and (b) mass distributions for primaries (thin lines) and members 
(bold lines) in the $\eta$ Cha cluster.  These distribution plots 
are flat beyond $r > 18$ arcmin, reflecting the absence of cluster 
members beyond this radius.  The dotted lines are the expected 
distributions for isothermal spheres with core radius $r_c = 18$ 
arcmin (0.5 pc).  The radial distribution (Figure 7a) of stars is 
consistent with an isothermal sphere.  However, the observed mass 
distribution (Figure 7b) appears more centrally concentrated than
the isothermal model, with the difference significant at the 
$P > 95$ percent confidence level for primaries ($P > 99$ 
percent for members) based on the 1-sample K-S test.

An alternative measure of mass segregation considers the radial
($U-V$) colour excess, where the photometric colour effectively 
acts as a mass surrogate.  \citet{Sagar89} studied 12 open 
clusters via integrated {\it UBV\,} colours using the observations 
of individual cluster members and demonstrated that clusters 
showing pronounced mass segregation also show significant radial 
variations in integrated colour, where $\Delta$($U-V$) = 
[($U-V$)$_{\rm outer~region}$ -- ($U-V$)$_{\rm inner~region}$)] 
$> 0.15$ mag.  For the $\eta$ Cha cluster stars, with no
available $U$-band photometry, we estimated the photospheric
$U$-band magnitude using the colour conversion tables of 
\citet{Kenyon95}\footnote{This is an appropriate technique even 
with the availability of $U$- and $B$-band fluxes, since PMS stars
often show $U$- and $B$-excesses due to chromospheric emission.}.
Comparing the integrated colour for members residing within
$r < 6$ arcmin (the inner region), and those outside (the outer
region) we find $\Delta$($U-V$) $\approx 2.0$ (ignoring $\eta$ 
Cha itself to reduce the stochastic effects of the 
brightest cluster member).  This colour excess for the $\eta$
Cha cluster exceeds all values found by \citet{Sagar89},
whose highest value was $\Delta$($U-V$) = 1.05 for NGC 6530.

From the radial mass and colour distributions, we conclude that 
significant mass segregation is present in the $\eta$ Cha cluster.  
This is the sparsest stellar cluster for which mass segregation 
has been seen.  Without detailed dynamical modeling and the
availability of precise space velocities for cluster members; 
see e.g. Delgado-Donate et al. (2003), our observations do not
as-yet allow us to distinguish between mass segregation arising 
from the cluster formation process or from dynamical two-body or 
$N$-body interactions.

A related question is whether the immediate environments of higher
mass stars have unusual densities of lower mass stars, e.g. Orion 
Trapezium OB stars show high numbers of nearby low-mass stars 
\citep{Preibisch99} which may support models where high mass stars 
form from the coalescence of many lower mass stars \citep{Testi98}. 
Two of the intermediate-mass stars, RS Cha (known double, and probable 
triple system) and HD 75505 (itself single), have $2-3$ low-mass 
stars residing within a few arcmin, while $\eta$ Cha itself (probable
binary) appears isolated.  We thus see no consistent excess nor
deficiency of lower mass stars in the close vicinity of the higher
mass stars.

\subsection{Cluster geometry}

There is a hint that the shape of the $\eta$ Cha cluster is 
flattened in Figure \ref{F1.fig} with a projected axis ratio 
of $\sim$ 2:1 in a NE-SW orientation, although the flattening 
is not statistically significant due to the small-$N$ population.  
Flattened-shapes have been documented in other young clusters, 
e.g. the Orion Nebula Cluster, Mon R2, and NGC 2024 
\citep{Hillenbrand98, Carpenter97, Lada91}.  These shapes may 
be the result of externally-triggered star formation, e.g. the 
passage of shock waves, or due to cloud-cloud collisions 
\citep{Clarke00}.  Possible internal triggers could involve an 
anisotropy of gravitational collapse due to magnetic fields, or 
possibly a non-spherical proto-cloud, e.g. \citet{Larson85}.

\subsection{Binary fraction}

The 17 stellar systems in the $\eta$ Cha cluster include 4 known 
binaries of which one is a probable triple, and 4 probable binaries, 
for a multiple fraction of $24-47$ percent. \citet{Kohler02} surveyed 
the RECX stars and the CTT star ECHA J0843.3-7905 for companions with
separations $< 8$ arcsec (800 AU) using $H$- and $K$-band speckle 
imaging.  They resolved only RECX 1AB and RECX 9B with separations 
of $\approx 20$ AU.  This indicates the other known and candidate 
binaries common to our studies (K\"ohler \& Petr-Gotzens did not 
observe the two new low-mass candidate binaries announced in this 
paper) have separations $< 13$ AU, the diffraction limit of the 3.5-m 
NTT at $K$-band. A possible constraint limiting the \citet{Kohler02} 
study is the $\Delta K \approx 2.5$ mag detection limit between 
primaries and companions.  However, with the exception of the probable 
low-mass companions to $\eta$ Cha and RS Cha AB, the other known and 
suspected binaries in the $\eta$ Cha cluster are near-equal mass 
(and therefore near-equal brightness) systems.  

The known/probable multiple fraction in the $\eta$ Cha cluster is 
$\sim 2-4 \times$ higher than the multiple fraction for solar-type
main-sequence stars in the Sun's neighborhood for separations 
$< 100$ AU \citep{Duquennoy91}.  For a group of 17 systems, the 
binary separation distribution of \citet{Duquennoy91} predicts 
$\sim 2.5$ multiple systems.

\citet{Kohler02} noted the non-detection of companions at separations 
$> 0.3$ arcsec or $> 30$ AU.  We confirm this result, for separations 
$> 1-2$ arcsec ($100-200$ AU), from our photometric survey which 
found no candidate cluster members located nearby known primaries.  
The distribution of \citet{Duquennoy91} predicts $\sim 2$ systems
with separations of $100-800$ AU.  As discussed by \citet{Kohler02},
this may indicate a deficiency of binaries with large separations in
the $\eta$ Cha cluster.

In addition, $3-6$ pairs among the 8 known/probable multiple systems 
of the $\eta$ Cha cluster have companions of near-equal mass (mass 
ratio $< 2:1$).   This result supports the notion that most binary 
stars form by fragmentation during the collapse of dense molecular 
cloud cores \citep{Bodenheimer00}.   Binaries resulting from star-disc
capture \citep{McDonald95} should show the companion mass distribution 
reflect almost random pairing from the IMF.

\citet{Kroupa95}, \citet{Bouvier97} and \citet{Mathieu00} suggest
that the density of the star-forming region and the cloud temperature 
can influence the binary fraction.   The high binary frequency in 
the $\eta$ Cha cluster is similar to the binary fraction in Taurus 
($43 \pm 5$ percent; K\"ohler \& Leinert 1998) which is $\sim 2 \times$ 
higher than that expected for field stars, and the binary fraction in 
Taurus-Auriga and Ophiuchus-Scorpius ($60 \pm 17$ percent;  Ghez 
et al. 1993), about $4 \times$ higher than that for field stars.  
In dense star-formation regions, e.g. the Trapezium \citep{Simon99} 
and NGC 2024, NGC 2068 and NGC 2071 \citep{Padgett97}, the binary 
fraction appears roughly equal to that of the field stars.  Together 
these consistent results support a relation between the binary fraction 
and the stellar density of star-forming regions.

\section{Conclusion Remarks}

An optical photometric survey of $45 \times 45$ arcmin 
($1.3 \times 1.3$ pc) centered on the $\eta$ Cha star cluster 
has added two new M5 -- M5.5 WTT stars to the known cluster 
population.  With no other cluster members found at $r < 23$ 
arcmin from the cluster centre, and as far as 32 arcmin in 
some directions, we have likely located all cluster primaries 
with spectral types earlier than $\sim$ M6 unless other stars 
have been more-distantly removed by evaporation or dynamical 
processes.  These two new stars contribute to a cluster 
population of 17 primaries ranging from spectral type 
B9 -- M5.5.  

We characterise the binary population of the cluster 
using a diversity of evidence ranging from light curve analysis, 
radial velocity variability, speckle imaging and colour-magnitude 
diagram placement to identify 4 confirmed and 5 probable stellar 
companions, most with low mass ratios ($1-2.5$:1).  The high-mass 
ratio binary population of the cluster remains largely undefined, 
except for probable companions to two of three early-type stars.
However, the known/likely multiple fraction of $24-47$ percent 
is already $2-4 \times$ higher than the observed field star 
multiplicity.

We investigate the $\eta$ Cha cluster IMF by comparison to the 
IMF of a well-characterised rich young cluster (Trapezium), and 
predict $20-29$ undiscovered low-mass stars and brown dwarfs 
($M = 0.025-0.15$ M$_{\odot}$) are likely to be associated 
with the known stellar population of 17 primaries ($21-26$
members).  We also compare the cluster IMF to that of other
sparse young stellar groups, and to the field IMF.  We find
no significant difference between the IMF of the $\eta$ Cha
cluster and the three comparator groups.  This suggests there
is no diversity in the IMF across several orders of magnitude
size scale of star formation in the Galaxy. 

From study of the cluster's spatial and mass distribution we find 
the $\eta$ Cha cluster has significant mass segregation, with 
$> 50$ percent of the stellar mass residing within the central 6 
arcmin (0.17 pc).  Since the $\eta$ Cha cluster is, at the same 
time, sparse (17 primaries), diffuse (average stellar density of 
$\approx 50$ stars\,pc$^{-3}$) and young (9 Myr), the cluster 
may be an excellent laboratory for distinguishing between mass 
segregation that is primordial in nature, or due to dynamical 
interaction processes.

\section*{Acknowledgments}  

ARL acknowledges the support of a UNSW@ADFA International
Postgraduate Research Scholarship.  WAL's research is supported
by UNSW@ADFA Faculty Research Grants and Special Research Grants.  
EDF thanks the University of New South Wales and Australian 
Defence Force Academy for hospitality.  LAC is supported by a 
National Research Foundation Postgraduate Scholarship.  We thank 
the SAAO and MSSSO time allocation committees for telescope time 
during 2002.  We also thank the referee for reminding us of the 
body of recent theoretical work underpinning cluster dynamics 
studies.

\bsp

\label{lastpage}

\end{document}